\documentclass{phbauth}
\usepackage{graphicx}
\usepackage{amssym}
\newcommand{\text}[1]{{\mathrm{#1}}}
\renewcommand{\vec}{\mathbf}

\begin{document}
\begin{frontmatter}
\title{Umklapp scattering at reconstructed quantum-Hall edges}

\author[tfp,ind]{U.~Z\"ulicke\thanksref{corresp}},
\author[ind]{A.~H.~MacDonald}

\address[tfp]{Institut f\"ur Theoretische Festk\"orperphysik,
              Universit\"at Karlsruhe, D-76128 Karlsruhe,
              Germany\thanksref{present}}
\address[ind]{Department of Physics, Indiana University,
              Bloomington, IN 47405, U.S.A.}

\thanks[corresp]{Corresponding author. Fax: +49 721 69 81 50;
e-mail: ulrich.zuelicke@phys.uni-karlsruhe.de.}
\thanks[present]{Present address.}

\date{15 June 1999}

\begin{abstract}
We study the low-lying excitations of a quantum-Hall sample that
has undergone {\em edge reconstruction\/} such that there exist three
branches of chiral edge excitations. Among the interaction processes
that involve electrons close to the three Fermi points is a new type
of {\em Umklapp\/}-scattering process which has not been discussed
before. Using bosonization and a refermionization technique, we obtain
exact results for electronic correlation functions and discuss the
effect Umklapp scattering has on the Luttinger-liquid properties of
quantum-Hall edges.
\end{abstract}
\end{frontmatter}

\section{Introduction}
The electronic structure at the edge of quantum-Hall (QH) systems
depends sensitively on the interplay between the external potential
confining the electrons to the finite sample, electrostatic
repulsion, as well as exchange and correlation effects. For an
ultimately sharp edge~\cite{sharpness}, a single branch of chiral
one-dimensional (1D) excitations is predicted to exist when the
filling factor $\nu=1/m$ where $m$ is an odd
integer~\cite{ahm:wen:90}. In that case, the dynamics of edge
excitations can be described~\cite{wen:rev} using a
Tomonaga-Luttinger (TL) model~\cite{tom:lutt} with only the
right-moving~\cite{rightmove} degrees of freedom present.
However, for a confining potential that is just not sharp enough to
stabilize a single-branch edge, a different configuration is
realized where a lump of electron charge is separated from the
bulk of the QH sample~\cite{ahm:aust:93,wen:prb:94}. Such a
{\em reconstructed\/} edge supports three branches of chiral 1D edge
excitations, two right-moving and one left-moving. For even weaker
confining potential, further reconstructions occur, leading to a
proliferation of edge-excitation modes~\cite{manymodes}. The
microscopic structure of a very smooth edge is dominantly determined
by electrostatics, which favors a phase separation of the 2D electron
system at the edge into a series of alternating compressible and
incompressible strips~\cite{smooth}.

Effective TL theories~\cite{wen:rev} describing single-branch and
multi-branch QH edges predict Luttinger-liquid behavior, i.e., power
laws governing the energy dependence of electronic correlation
functions. The characteristic exponents of these power laws depend,
in general, on details of the microscopic edge structure. However, in
the absence of coupling between different chiral edge branches or, in
some cases, due to disorder effects~\cite{mpaf:prb:95a}, power-law
exponents turn out to be universally dependent on the bulk filling
factor. At present, microscopic details of the edge structure
that is realized in experiment~\cite{amc:prl} are not fully known. To
facilitate a realistic comparison between theory and experiment, it
is necessary to study the low-lying edge excitations of reconstructed
and smooth edges and investigate interaction effects on the
Luttinger-liquid power-law exponents when more than one branch of
edge excitations is present.

\section{Derivation of the effective edge theory}
We focus on the edge of a spin-polarized~\cite{spinapology} QH sample
at $\nu=1$ that has undergone reconstruction such that three branches
of edge excitations are present. To be specific, we choose the Landau
gauge where lowest-Landau-level (LLL) basis states $\chi_k(x,y) =
\Phi_k(y)\,\exp\{ikx\}/\sqrt{L}$ are labeled by a 1D wave vector $k$.
Here, $\ell=\sqrt{\hbar c/|e B|}$ denotes the magnetic length, $L$ is
the edge perimeter, and $\Phi_k(y) = \exp\{-(y - k\ell^2)^2/(2\ell^2)
\}/\sqrt{\pi^{1/2}\ell}$. In the absence of interactions between
different edge branches, the ground state would be a generalized
Fermi-sea state that is a Slater determinant of LLL basis states
whose wave-vector label satisfies $k\le k_{\text{F}}^{\text{(R)}}$
or $k_{\text{F}}^{\text{(W)}}\le k \le k_{\text{F}}^{\text{(B)}}$.
The Fermi `surface' consists of three (Fermi) points
$k_{\text{F}}^{\text{(R)}} < k_{\text{F}}^{\text{(W)}}<
k_{\text{F}}^{\text{(B)}}$. As in Tomonaga's approach to
interacting 1D electron systems~\cite{tom:lutt}, long-wave-length
electronic excitations at the reconstructed edge can be identified
according to which Fermi point they belong to. This makes it
possible to rewrite the long-wave-length part of the electron
operator as follows:
\begin{eqnarray}\label{electron}
\psi(\vec r) &=& \Phi_{k^{\text{(R)}}_{\text{F}}}(y)\,
e^{i k^{\text{(R)}}_{\text{F}} x}\, \psi^{\text{(R)}}(x) \nonumber
\\ && +
\Phi_{k^{\text{(W)}}_{\text{F}}}(y)\, e^{i k^{\text{(W)}}_{\text{F}} x}
\, \psi^{\text{(W)}}(x) \nonumber \\ && +
\Phi_{k^{\text{(B)}}_{\text{F}}}(y)\, e^{i
k^{\text{(B)}}_{\text{F}} x}\, \psi^{\text{(B)}}(x)\quad .
\end{eqnarray}
Here, $\vec r = (x, y)$ denotes the coordinate vector in the 2D plane,
and the operator $\psi^{\text{(R,W,B)}}(x)$ creates an electron
belonging to the chiral 1D edge branch labeled R, W, B, respectively.
The interaction part of the 2D Hamiltonian for electrons in the LLL is
\begin{equation}\label{intham}
H_{\mathrm{int}} = \frac{1}{2} \int\!\!\int d^2 r\, d^2 r^\prime \,\,
V(\vec r - \vec{r^\prime})\,\varrho(\vec r)\,\varrho(\vec{r^\prime})
\quad ,
\end{equation}
where $\varrho(\vec r)= \psi^\dagger(\vec r)\psi(\vec r)$ is the
electron density. We consider the case when electrons interact via
unscreened Coulomb interaction,
\begin{equation}
V(\vec r - \vec{r^\prime}) = \frac{e^2/\epsilon}{\sqrt{(x - 
x^\prime)^2 + (y - y^\prime)^2}}\quad .
\end{equation}
We obtain the low-energy part of $H_{\mathrm{int}}$ by inserting
Eq.~(\ref{electron}) into Eq.~(\ref{intham}); it is effectively 1D
and comprises various scattering processes of electrons that are close
to one of the three Fermi points. Terms corresponding to forward
scattering and backscattering~\cite{backscatt} have been discussed
before~\cite{wen:prb:94}. Together with the one-body part of the
original 2D Hamiltonian, they can be rewritten in terms of a TL model
Hamiltonian, $H_{\mathrm{TL}}$, which is quadratic in the Fourier
components $\varrho_q^{(\alpha)}$ of the chiral 1D densities $\big[
\psi^{(\alpha)}(x)\big]^\dagger\psi^{(\alpha)}(x)$ (here, $\alpha\in\{
\text{R,W,B}\}$). In the long-wave-length limit, where Coulomb matrix
elements dominate the bare Fermi velocities, the three normal modes of
$H_{\mathrm{TL}}$ are~\cite{wen:prb:94} a)~the edge-magnetoplasmon
mode, $\varrho^{\text{(emp)}} = \varrho^{\text{(B)}} +
\varrho^{\text{(R)}} + \varrho^{\text{(W)}}$, which is right-moving,
and b)~two linearly dispersing neutral modes, a right-moving one given
by $\varrho^{\text{(rn)}}=(\varrho^{\text{(B)}} - \varrho^{\text{(R)}}
)/\sqrt{2}$, and the left-moving neutral mode $\varrho^{\text{(ln)}}=
(\varrho^{\text{(B)}} + \varrho^{\text{(R)}} + 2
\varrho^{\text{(W)}})/\sqrt{2}$.

\section{Umklapp scattering}
\begin{figure}[t]
\begin{center}\leavevmode
{\includegraphics[width=0.78\linewidth]{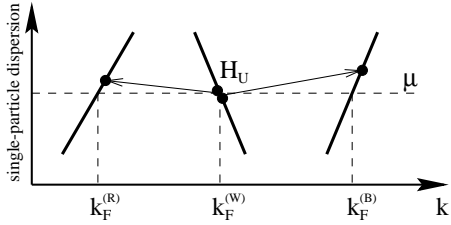}}
\caption{Schematic depiction of Umklapp scattering at a reconstructed
QH edge. We linearized the non-interacting single-particle dispersion
close to intersection points with the chemical potential $\mu$ that
define the three Fermi points. Shown is one interaction process
represented by $H_{\text{U}}$ where two left-moving electrons from the
W-branch are scattered into the R and B-branches.\vspace{-.2truecm}}
\label{umfig}\end{center}\end{figure}

In addition to forward and backscattering, the following term occurs
in the effective 1D Hamiltonian describing the low-energy excitations
of a reconstructed QH edge:
\begin{eqnarray}
&&H_{\mathrm U} = \int\!\!\int dx\,\, dx^\prime\,\, V_{\text{U}}(x-
x^\prime)\bigg\{\big[\psi^{\text{(R)}}(x)\big]^\dagger\big[
\psi^{\text{(B)}}(x^\prime)\big]^\dagger\nonumber\\ && \times
\psi^{\text{(W)}}(x^\prime)\,\psi^{\text{(W)}}(x)
e^{i D \frac{x-x^\prime}{2} - i\delta
\frac{x+x^\prime}{2}} + {\mathrm{H.c.}} \bigg\}\, .
\end{eqnarray}
Here we introduced the parameters $\delta=k_{\text{F}}^{\text{(B)}} +
k_{\text{F}}^{\text{(R)}} - 2 k_{\text{F}}^{\text{(W)}}$ and $D=
k_{\text{F}}^{\text{(B)}}-k_{\text{F}}^{\text{(R)}}$. The distance
$D \ell^2$ corresponds to the width of the edge. Note that
$H_{\text{U}}$ represents interaction processes (and their
time-reversed version) where two electrons from the left-moving
W-branch are scattering off each other such that one of them ends up
in the right-moving R-branch and the other one in the right-moving
B-branch. (See Fig.~\ref{umfig}.) Interaction processes converting two
left-movers into two right-movers (and {\it vice versa\/}) are
familiar from lattice models for conventional interacting 1D electron
systems; there they are called {\em Umklapp\/} processes~\cite{umlit}.
Based on that analogy, we adopt the term {\em Umklapp scattering\/}
for the interaction processes given by $H_{\text{U}}$. Note that
momentum conservation implies a commensuration issue for Umklapp
scattering. In the 1D Hubbard model, low-energy properties are only
affected by Umklapp processes if the large momentum transfer accrued
($4 k_{\text{F}}$) is close to a reciprocal-lattice
vector~\cite{umlit} which is the case, e.g., at half-filling.
Similarly, Umklapp scattering at a reconstructed edge is most relevant
in the symmetric case when $\delta=0$. Note also that the matrix
element $V_{\text{U}}(x-x^\prime)$ gets small rapidly with increasing
$D$ and $\delta$; it is given by
\begin{eqnarray}
&& V_{\text{U}}(x-x^\prime) = \frac{e^2}{\epsilon\ell}\,\,\frac{\exp
\left\{ -\frac{\ell^2}{8}\left[\delta^2 + D^2\right]\right\}}{\sqrt{2
\pi}}\nonumber \\ && \times\int d\kappa \,\, \frac{\exp\left\{
-\frac{1}{2} \left[\kappa - \ell D/2\right]^2\right\}}{\sqrt{(x -
x^\prime)^2/\ell^2 + \kappa^2}} \, .
\end{eqnarray}

\section{Bosonization}
The new Umklapp process does not conserve particle number in each
edge branch separately. Therefore, $H_{\text{U}}$ cannot be
written in terms of a Tomonaga-Luttinger model. However, using the
bosonization identity\cite{bosid} for the 1D fermionic operators,
\begin{eqnarray}
\psi^{\text{(R)}}(x) &=& 1/\sqrt{L}\,\,\, {\mathbf :}\,\exp [ i
\,\phi^{\text{(R)}}(x)] \, {\mathbf :} \,\, ,\\
\psi^{\text{(W)}}(x) &=& 1/\sqrt{L}\,\,\, {\mathbf :}\,\exp [ -i
\, \phi^{\text{(W)}}(x)] \, {\mathbf :}\,\, ,\\
\psi^{\text{(B)}}(x) &=& 1/\sqrt{L}\,\,\, {\mathbf :}\,\exp [ i
\, \phi^{\text{(B)}}(x)] \, {\mathbf :}\,\, ,
\end{eqnarray}
where ${\mathbf :}\dots{\mathbf :}$ symbolizes normal ordering, and
\begin{equation}\label{phifields}
\phi^{(\alpha)}(x) = i\, \frac{2\pi}{L} \sum_{q\ne 0}
\frac{e^{- i q x}}{q}\,\, \varrho_q^{(\alpha)}\quad ,
\end{equation}
it is possible to rewrite $H_{\text{U}}$ entirely in terms of
bosonic degrees of freedom:
\begin{eqnarray}\label{bosum}
H_{\text{U}} &=& 2 \Lambda^2 g_{\text{U}} \int d x \, \cos\big[
\phi^{\text{(ln)}}(x) + \delta x\big]\, , \\
\phi^{\text{(ln)}}(x) &=& \phi^{\text{(R)}}(x) +\phi^{\text{(B)}}(x)
+ 2\phi^{\text{(W)}}(x)\, .
\end{eqnarray}
Expressions of the form~(\ref{bosum}) are routinely obtained within
the bosonized description of Umklapp scattering in conventional 1D
electron systems~\cite{umlit}. The parameter $\Lambda$ is a
physical ultraviolet cut-off; for the situation considered in this
work, we have $\Lambda<\sim (D\ell^2)^{-1}$. The effective Umklapp
coupling constant, $g_{\text{U}}$, is derived from the original
Coulomb interaction between the electrons. It is given by
\begin{eqnarray}
&&g_{\text{U}} = \frac{e^2}{\epsilon}\,\exp\left\{-\frac{\ell^2[\delta^2
+ D^2]}{8}\right\}\,\sqrt{\frac{2}{\pi}} \nonumber \\ &&\times
\int_{\infty}^{\infty}
d\kappa \, e^{-\frac{1}{2}\left[\kappa - \ell \frac{D}{2}\right]^2}\,
\int_{0}^{\infty} d\eta \,\, \frac{\sin\left[\ell D \eta/2 \right]}
{ \sqrt{\eta^2 + \kappa^2}}\, .
\end{eqnarray}
Note that $\phi^{\text{(ln)}}(x)$ is a chiral boson field given in
terms of the Fourier components $\varrho_q^{\text{(ln)}}$ as expressed
in Eq.~(\ref{phifields}). The Umklapp part of the Hamiltonian,
$H_{\text{U}}$, introduces a self-interaction of the left-moving
neutral normal mode of $H_{\text{TL}}$. The edge-magnetoplasmon mode
and the right-moving neutral mode are unaffected by Umklapp scattering
and remain free.

\section{Exact solution: Summary of results}
We have been able to solve the theory including Umklapp {\em exactly}
for arbitrary $\delta$ and $g_{\text{U}}$ using a refermionization
technique~\cite{tocome} whereby the Hamiltonian for the chiral 1D
bosonic field $\phi^{\text{(ln)}}(x)$ with interaction $H_{\text{U}}$
is mapped onto that of a chiral 1D pseudo-spin-1/2 fermion in an
external magnetic field that is perpendicular to the pseudo-spin
quantization axis. In addition to the characteristic energy scale for
Umklapp scattering, given by $\Delta_0=2\Lambda g_{\text{U}}$, the
parameter $\xi=\hbar v_{\text{ln}}|\delta|/(2 \Lambda g_{\text{U}})$
emerges from the calculation that measures the ineffectiveness of
Umklapp scattering due to deviation from perfect commensuration. We
obtain the spectral functions for tunneling into the chiral R,W,B edge
branches. Due to Umklapp scattering, a crossover occurs in their
energy dependence between different power laws. With $\alpha\in\{
\text{R,W,B}\}$, we find
\begin{equation}
{\mathcal A}^{(\alpha)}(\varepsilon)\propto \left\{
\begin{array}{cl}
\varepsilon^{[\lambda^{(\alpha)}_{\text{emp}}]^2 + [\lambda^{(
\alpha)}_{\text{rn}}]^2-1} & \mbox{for } \varepsilon < \Delta \\
\varepsilon^{[\lambda^{(\alpha)}_{\text{emp}}]^2 +
[\lambda^{(\alpha)}_{\text{rn}}]^2 + [\lambda^{(\alpha
)}_{\text{ln}}]^2-1} & \mbox{for } \varepsilon > \Delta \end{array}
\right.\, ,
\end{equation}
where the crossover energy scale is $\Delta=\Delta_0\big[\sqrt{1+
\xi^2} - \xi\big]$, and the $\lambda^{(\alpha)}_{\beta}$ are the
Bogoliubov coefficients relating the chiral density fluctuations
localized at the R,W,B branches to the normal modes of
$H_{\text{TL}}$. Based on a realistic model for a reconstructed edge,
we obtained the order-of-magnitude estimate $\Delta\sim 10
\dots 100\,\mu$eV. As the central result of our study, we
find that Umklapp scattering {\em diminishes\/} the value of the
power-law exponent in the tunneling density of states below the 
characteristic energy scale $\Delta$.

\section*{Acknowledgments}
This work was funded in part by NSF Grant No.\ DMR-9714055 and
Sonderforschungs\-bereich 195 der Deutschen Forschungsgemeinschaft.
U.Z.\ acknowledges helpful discussions with I.~Affleck, W.~Apel,
C.~de~C.~Chamon, E.~Fradkin, S.~M.~Girvin, B.~I.~Halperin,
V.~Meden, N.~P.~Sandler, and J.~von~Delft.


\begin{thebibliography}{10}
\bibitem{sharpness}
The sharpness/smoothness of the edge can be quantified by the slope
of the bare confining potential at the sample boundary.
\bibitem{spinapology}
We only consider the case when the spin-degree of freedom is frozen
out such that it is appropriate to assume electrons to be spinless.
\bibitem{backscatt}
For spinless electrons, backscattering terms can be accounted for
within a TL model because they only renormalize forward-scattering
amplitudes.
\end{thebibliography}
\end{document}